\documentclass{elsart}

\usepackage{adnd}
\usepackage{longtable}

\usepackage{mathptmx}

\usepackage{amsmath}

\usepackage{amssymb,amstext}
\usepackage[pdftex,
        letterpaper=true,
        hyperindex=true,
        breaklinks=true,
        colorlinks=false,
        citecolor=blue,
        pdftitle={},
        pdfauthor={}]
{hyperref}
\usepackage{graphicx}

\begin{document}

\begin{frontmatter}

\journal{Atomic Data and Nuclear Data Tables}

\copyrightholder{Elsevier Science}

\runtitle{Cadmium}
\runauthor{Amos}

%% Author, fill in article title here

\title{Discovery of the Cadmium Isotopes}

%% Fill in author list here

\author{S. Amos}
\and
\author{M.~Thoennessen\corauthref{cor}}\corauth[cor]{Corresponding author.}\ead{thoennessen@nscl.msu.edu}

\address{National Superconducting Cyclotron Laboratory and \\ Department of Physics and Astronomy, Michigan State University, \\East Lansing, MI 48824, USA}

\date{September 3, 2009} %please do not use \today, use actual date of version

\begin{abstract}
 Thirty-seven cadmium isotopes have so far been observed; the discovery of these isotopes is discussed.  For each isotope a brief summary of the first refereed publication, including the production and identification method, is presented.
\end{abstract}

\end{frontmatter}

%%% Keywords and subject classification are not used in ADNDT
%%%\begin{keywords}
%%%Insert list of keywords here.
%%%\end{keywords}

%%%\begin{subject}[Insert header for classifications]
%%%Use only if your journal has a subject classification requirement
%%%\end{subject}

%%% The table of contents should start a new page. This command will
%%% automatically pull all the unstarred \section, \subsection and
%%% \subsubsection titles into the Contents. Starred versions need to be
%%% done manually using
%%%            \addcontentsline{toc}{[[sub]sub]section}{Section title}
%%% at the correct place. Examples are given below.

%%% The lists of data figures and data tables are created automatically
%%% by the \listofDfigures and \listofDtables commands.

\newpage
\tableofcontents
%%\listofDfigures
\listofDtables

\vskip5pc

\section{Introduction}\label{s:intro}

The discovery of the cadmium isotopes is discussed as part of the series of the discovery of isotopes which began with the cerium isotopes in 2009
\cite{Gin09}. The purpose of this series is to document and summarize the discovery of the isotopes. Guidelines for assigning credit for discovery are (1) clear identification, either through decay-curves and relationships to other known isotopes, particle or $\gamma$-ray spectra, or unique mass and Z-identification, and (2) publication of the discovery in a refereed journal. The authors and year of the first publication, the laboratory where the isotopes were produced as well as the production and identification methods are discussed. When appropriate, references to conference proceedings, internal reports, and theses are included. When a discovery includes a half-life measurement the measured value is compared to the currently adapted value taken from the NUBASE evaluation \cite{Aud03} which is based on the ENSDF database \cite{ENS08}.

\section{Discovery of $^{96-132}$Cd}

Thirty-seven cadmium isotopes from A = $96-132$ have been discovered so far; these include 8 stable, 12 proton-rich and 17 neutron-rich isotopes.  According to the HFB-14 model \cite{Gor07}, $^{163}$Cd should be the last particle stable neutron-rich nucleus ($^{159}$Cd is calculated to be unbound). Along the proton dripline five more isotopes are predicted to be stable and it is estimated that three additional nuclei beyond the proton dripline could live long enough to be observed \cite{Tho04}. Thus, there remain 38 isotopes to be discovered. About 50\% of all possible cadmium isotopes have been produced and identified so far.

\begin{figure}
	\centering
	\includegraphics[scale=.5]{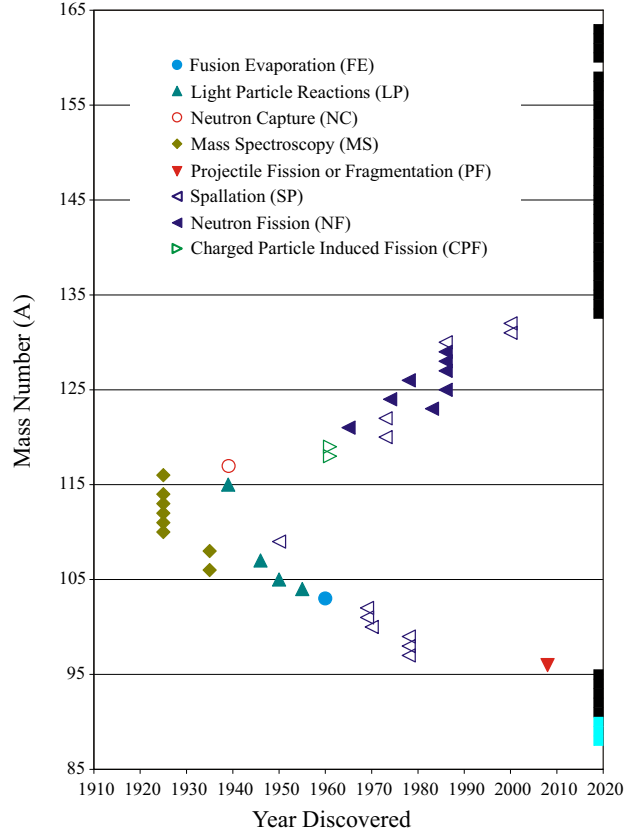}
	\caption{Cadmium isotopes as a function of time they were discovered. The different production methods are indicated. The solid black squares on the right hand side of the plot are isotopes predicted to be bound by the HFB-14 model.  On the proton-rich side the light blue squares correspond to unbound isotopes predicted to have lifetimes larger than $\sim 10^{-9}$~s.}
\label{f:year}
\end{figure}

Figure \ref{f:year} summarizes the year of first discovery for all cadmium isotopes identified by the method of discovery.  The range of isotopes predicted to exist is indicated on the right side of the figure.  The radioactive cadmium isotopes were produced using fusion evaporation (FE), projectile fragmentation or projectile fission (PF), light-particle reactions (LP), neutron capture (NC), neutron-induced fission (NF), charged-particle induced fission (CPF) and spallation reactions (SP). The stable isotopes were identified using mass spectroscopy (MS). Heavy ions are all nuclei with an atomic mass larger than A=4 \cite{Gru77}. Light particles also include neutrons produced by accelerators. In the following, the discovery of each cadmium isotope is discussed in detail.

\subsection*{$^{96}$Cd}\vspace{-0.85cm}

In 2008 Bazin \textit{et al.} reported the discovery of $^{96}$Cd in \textit{Production and $\beta$ Decay of rp-Process Nuclei $^{96}$Cd, $^{98}$In, and $^{100}$Sn} \cite{Baz08}. At the National Superconducting Cyclotron Laboratory at Michigan State University a 120 MeV/nucleon $^{112}$Sn primary beam reacted with a 195 mg/cm$^{2}$ $^{9}$Be target and produced $^{96}$Cd, $^{98}$In, and $^{100}$Sn. The A1900 fragment separator and the Radio Frequency Fragment Separator were then used to purify the products which were identified by energy loss, time-of-flight measurements, and $gamma$-ray tagging. The halflives of the isotopes were measured after implentation in the NSCL beta counting system. ``The half-life of $^{96}$Cd, which was the last experimentally unkown waiting point half-life of the astrophysical {\it rp} process, is  1.03$^{+0.24}_{-0.21}$~s.'' This half-life measurement is the only value for $^{96}$Cd at this time.

\subsection*{$^{97-99}$Cd}\vspace{-0.85cm}

Three new isotopes, $^{97}$Cd, $^{98}$Cd, and $^{99}$Cd, were discussed in the 1978 publication of \textit{Beta-Delayed Proton Emission from $^{97}$Cd and $^{99}$Cd} by Elmroth \textit{et al.} \cite{Elm78}. A 115 g/cm$^{2}$ natural tin target was bombarded by 600~MeV protons from CERN's synchro-cyclotron forming $^{97}$Cd, $^{98}$Cd, and $^{99}$Cd in spallation reactions. The isotopes were separated and identified with the ISOLDE electromagnetic isotope separator. ``Two new beta-delayed proton precursors, $^{97}$Cd and $^{99}$Cd, have been identified at the ISOLDE on-line isotope separator... A search for $^{98}$Cd was also performed and it was found to be a pure $\beta$-emitter with a probable half-life of $\sim$ 8 sec.'' The half-life of $^{99}$Cd was measured to be 16(3)~s. The $^{99}$Cd half-life corresponds to the currently accepted value, and the $^{98}$Cd half-life is close to the accepted value of 9.2(3)~s.

\subsection*{$^{100}$Cd}\vspace{-0.85cm}

Hnatowich \textit{et al.} identified $^{100}$Cd for the first time in 1970 as reported in \textit{The decay of Cadmium isotopes of mass 100, 101, and 102 to isomers in silver} \cite{Hna70}. A high purity molten tin target was irradiated with 600 MeV protons from the CERN 600 MeV synchro-cyclotron and $^{100}$Cd was produced in the Sn(p,3pxn) spallation reaction. The isotopes were separated and identified with the ISOLDE electromagnetic isotope separator. ``We have measured the half-life of the new nuclide $^{100}$Cd by beta counting of on-line sources in a proportional counter. From these observations, our value for the $^{100}$Cd half-life, with an estimated error, is 1.1$\pm$0.3 min.'' This half-life value is close to the currently accepted value of 49.1(5)~s.

\subsection*{$^{101,102}$Cd}\vspace{-0.85cm}

Hansen \textit{et al.} reported the first observation of $^{101}$Cd and $^{102}$Cd in the paper \textit{Decay Characteristics of Short-Lived Radio-Nuclides Studied by On-Line Isotope Separator Techniques} in 1969 \cite{Han69}. Protons of 600~MeV from the CERN synchrocyclotron bombarded a molten tin target and cadmium was separated using the ISOLDE facility. The paper summarized the ISOLDE program and did not contain details about the individual nuclei other than in tabular form. The measured half-lives of $^{101}$Cd and $^{102}$Cd were 1.2(2)~m, and 5.5(5)~m respectively. The measurement for $^{101}$Cd is included in the weighted average of the currently accepted half-life of  1.36(5)~m, while the measurement for $^{102}$Cd is the currently accepted value. A previous report of half-life measurements of 15~m and 30~m for $^{101}$Cd and $^{102}$Cd, respectively \cite{But66}, could not be confirmed.

\subsection*{$^{103}$Cd}\vspace{-0.85cm}

Preiss \textit{et al.} published the discovery of $^{103}$Cd in 1960 in \textit{A New Isotope: Cadmium-103} \cite{Pre60}. A 160~MeV $^{16}$O beam from the heavy-ion accelerator of the Sterling Chemistry Laboratory at Yale University bombarded molybdenum oxide targets and formed $^{103}$Cd in the fusion-evaporation reactions $^{92-100}$Mo($^{16}$O,2p3-11n). $^{103}$Cd was identified by its decay curve and $\gamma$-ray measurements following chemical separation. ``A new activity, produced by 160 MeV O$^{16}$ bombardments of thick molybdenum oxide targets, is assigned to Cd$^{103}$ on the basis of chemistry and its decay to Ag$^{103}$.'' The measured half-life of 10.0(15)~m is close to the currently accepted value of 7.3(1)~m.

\subsection*{$^{104}$Cd}\vspace{-0.85cm}

In \textit{The New Isotopes Cd$^{104}$ and Ag$^{104}$}, Johnson reported the observation of $^{104}$Cd in 1955 \cite{Joh55}. Protons, accelerated to 50 MeV by the McGill University 100 MeV synchrocyclotron, bombarded metallic silver. $^{104}$Cd was studied with a 180-degree spectrograph, a lens spectrometer, and a scintillation spectrometer. ``Two new neutron deficient isotopes Cd$^{104}$ (59 min.) and Ag.$^{104}$ (27 min) have been produced by the reaction Ag$^{107}$(p,4n)Cd$^{104}$ at 50 Mev. in the McGill University synchrocyclotron and by the subsequent growth of silver from the cadmium.'' The measured half-life of 59~m is in agreement with the 57.7(10)~m accepted value.

\subsection*{$^{105}$Cd}\vspace{-0.85cm}

The new isotope $^{105}$Cd was discussed in \textit{Radioactive Isotopes of Ag and Cd} by Gum and Pool in 1950 \cite{Gum50}. Palladium was bombarded with 20 MeV alpha particles at Ohio State University producing $^{105}$Cd in the reaction ($\alpha$,n). ``A Cd activity with a 57$\pm$2-min. half-life was produced when Pd was activated with 20-Mev alpha-particles... Cd bombarded with fast neutrons also yielded this activity. However, the activity is not observed when Ag is bombarded with deuterons or when Cd is bombarded with slow neutrons. These observations lead to the mass assignment of 105 for the 57-min half-life activity.'' This half-life agrees with the currently accepted value of 55.5(4)~m.

\subsection*{$^{106}$Cd}\vspace{-0.85cm}

Aston described the discovery of $^{106}$Cd in the 1935 article \textit{The Isotopic Constitution and Atomic Weights of Hafnium, Thorium, Rhodium, Titanium, Zirconium, Calcium, Gallium, Silver, Carbon, Nickel, Cadmium, Iron and Indium} \cite{Ast35}. Cadmium methyl was analyzed by anode rays using ordinary gas discharge. ``As was hoped, the new spectra revealed three more isotopes 106, 108, 115, and enabled photometric estimates of the constitution of this very complex element...''

\subsection*{$^{107}$Cd}\vspace{-0.85cm}

In 1946 $^{107}$Cd was correctly identified in the article, \textit{Isotopic Assignment of Cd and Ag Activities}, by Helmholz \cite{Hel46}. A cadmium target enriched in $^{106}$Cd was bombarded by slow neutrons produced in the Berkeley 60 inch cyclotron. The isotope was identified by absorption curves of the emitted electrons. ``The 6.7 hour period was the predominant activity in the 106 sample after a few hours bombardment... The results then assign the 6.7 hr. Cd to Cd$^{107}$,...'' This half-life agrees with the accepted value of 6.50(2)~h. A 6.7~h half-life had been observed earlier, but no mass assignment could be made \cite{Del39,Alv40}. Kirshnan and Gant assigned a 6.8~h half-life to either an isomeric state of $^{108}$Cd or $^{110}$Cd or to radioactive $^{107}$Cd or $^{109}$Cd \cite{Kri39}. Several other measurements were performed which could not make a unique mass assignment and referred to the 6.7~h activity as either $^{107}$Cd or $^{109}$Cd \cite{Bra45a,Bra45b,Bra46}.

\subsection*{$^{108}$Cd}\vspace{-0.85cm}

Aston described the discovery of $^{108}$Cd in the 1935 article \textit{The Isotopic Constitution and Atomic Weights of Hafnium, Thorium, Rhodium, Titanium, Zirconium, Calcium, Gallium, Silver, Carbon, Nickel, Cadmium, Iron and Indium} \cite{Ast35}. Cadmium methyl was analyzed by anode rays using ordinary gas discharge. ``As was hoped, the new spectra revealed three more isotopes 106, 108, 115, and enabled photometric estimates of the constitution of this very complex element...''

\subsection*{$^{109}$Cd}\vspace{-0.85cm}

In \textit{The Spallation Products of Antimony Irradiated with High Energy Particles} Lindner and Perlman discuss their accurate identification of $^{109}$Cd in 1950 \cite{Lin50}. $^{109}$Cd was produced by bombarding a thick antimony target with deuterons and $\alpha$-particles from the Berkeley 134-inch cyclotron with energies of 190~MeV and 380~MeV, respectively. Decay curves were recorded and absorption measurements were performed following chemical separation.  ``This [The 330-day Cd$^{109}$] activity was apparent in the decay curve taken through beryllium soon after the 6.7-hr. Cd$^{107}$ had disappeared. Ultimately it could be seen in the curve taken without absorber after the 42-day Cd$^{115}$ had decayed. Identification was made certain by removal of its daughter, 40-sec. Ag$^{109m}$.'' The half-life of 330 days is near the currently accepted half-life of 461.4(12)~d. In 1940 Krishnan had reported a half-life of approximately one year but could not determine whether it was $^{109}$Cd or $^{107}$Cd \cite{Kri40}. In 1945 Bradt {\it et al.} measured a half-life of 158(7)~d and assigned it to either $^{109}$Cd or $^{107}$Cd \cite{Bra45b}. A year later they corrected the half-life to 330~days \cite{Bra46}. At the same time Helmholz had assigned the 158~d activity to $^{109}$Cd, however, we do not credit him with the discovery because the half-life differs significantly from the correct value. The work by Lindner and Perlman was submitted in February 1950 and it should be mentioned that in May of 1950 Cork {\it et al.} reported a half-life of 250~d for $^{109}$Cd \cite{Cor50} and in June of 1950 Gum and Pool measured a half-life of 470~days \cite{Gum50}.

\subsection*{$^{110-114}$Cd}\vspace{-0.85cm}

In 1925 Aston reported the first observation of $^{110-114}$Cd in \textit{The Mass Spectra of Chemical Elements, Part VI. Accelerated Anode Rays Continued} \cite{Ast25}. The mass spectra were measured at the Cavendish Laboratory in Cambridge, UK using an anode ray discharge tube containing cadmium fluoride and lithium fluoride. ``This element, unsuccessfully attacked before, has now yielded to improved experimental conditions... Cadmium has six isotopes 110, 111, 112, 113, 114, 116.'' These isotopes were all assumed to be stable, however, $^{113}$Cd decays with a half-life of $7.7(3)\cdot10^{15}$~y.

\subsection*{$^{115}$Cd}\vspace{-0.85cm}

The isotope $^{115}$Cd was first correctly identified by Goldhaber {\it et al.} in 1939 in \textit{Radioactivity Induced by Nuclear Excitation} \cite{Gol39}. Cadmium was irradiated with neutrons produced by bombarding lithium with deuterons of 950 KeV at the Cavendish Laboratory. $^{115}$Cd was produced in the reaction $^{116}$Cd(n,2n). Decay curves were measured with a Geiger-M\"uller counter following chemical separation. ``Our experiments show that In$^{115*}$ grows from the parent Cd$^{115}$ which decays with a half-life time of 2.5 days.''  This measurement agrees with the accepted half-life of 53.46(10)~h. In 1937, Cork and Thornton had reported half-lives of 58~h and 4.3~h and incorrectly assigned them to $^{117}$Cd and $^{115}$Cd, respectively \cite{Cor37}. A few months later Mitchell \cite{Mit37} measured half-lives of 52~h and 5~h, agreeing with the assignment by Cork and Thornton. Seven months after the submission of the Goldhaber paper, Cork and Lawson independently reversed their assignment of newly measured half-lives of 56~h and 3.75~h to $^{115}$Cd and $^{117}$Cd, respectively \cite{Cor39}. It should be noted that Aston had incorrectly reported the observation of stable $^{115}$Cd in 1935 \cite{Ast35}.

\subsection*{$^{116}$Cd}\vspace{-0.85cm}

In 1925 Aston reported the first observation of $^{116}$Cd in \textit{The Mass Spectra of Chemical Elements, Part VI. Accelerated Anode Rays Continued} \cite{Ast25}. The mass spectra were measured at the Cavendish Laboratory in Cambridge, UK using an anode ray discharge tube containing cadmium fluoride and lithium fluoride. ``This element, unsuccessfully attacked before, has now yielded to improved experimental conditions... Cadmium has six isotopes 110, 111, 112, 113, 114, 116.''

\subsection*{$^{117}$Cd}\vspace{-0.85cm}

The isotope $^{117}$Cd was first correctly identified by Goldhaber {\it et al.} in 1939 in \textit{Radioactivity Induced by Nuclear Excitation} \cite{Gol39}. Cadmium was irradiated with neutrons produced by bombarding lithium with deuterons of 950 KeV at the Cavendish Laboratory. Decay curves were measured with a Geiger-M\"uller counter following chemical separation. The observation of $^{117}$Cd was only mentioned in a footnote: ``They [Cork and Thornton \cite{Cor37}] also reported a radioactive isotope of In of 2.3-hr. half-life time which they ascribed to In$^{117}$ and which appeared to grow from a 58-hr. Cd parent. We find this indium isotope only when we precipitate indium from the cadmium solution in a time less than two days after irradiation. By successive separations we find that it grows from a 4-hr. Cd parent, a radioactive isotope which has been previously reported by Cork and Thornton and at the time assumed to be Cd$^{115}$.'' This half-life was assigned to $^{115}$Cd in a table and must have been produced by neutron capture on $^{116}$Cd. The 4~h half-life is in reasonable agreement with the presently accepted value of 2.49(4)~h. As mentioned by Goldhaber {\it et al.}, Cork and Thornton in 1937 had reported half-lives of 58~h and 4.3~h and incorrectly assigned them to $^{117}$Cd and $^{115}$Cd, respectively \cite{Cor37}. A few months later Mitchell \cite{Mit37} measured half-lives of 52~h and 5~h, agreeing with the assignment by Cork and Thornton. Seven months after the submission of the Goldhaber paper, Cork and Lawson independently reversed their assignment of newly measured half-lives of 56~h and 3.75~h to $^{115}$Cd and $^{117}$Cd, respectively \cite{Cor39}.

\subsection*{$^{118}$Cd}\vspace{-0.85cm}

 The 1961 article \textit{Decay of 49-min Cd$^{118}$ and 5.1-sec In$^{118}$} by Gleit and Coryell reported the discovery of $^{118}$Cd \cite{Gle61a}. Deuterons accelerated to 14 MeV by the M.I.T. cyclotron bombarded natural uranium foils. $^{118}$Cd was produced by fission and identified by its $\beta$-decay following chemical separation. ``An average of nine samples, each followed for more than seven half-lives, yields a half-life of 49.0$\pm$1.5~min for Cd$^{118}$.'' This half-life agrees with the currently accepted value of 50.3(2)~m. A half-life of 30~m had been reported in 1953 in a conference proceedings \cite{Cor53}.

\subsection*{$^{119}$Cd}\vspace{-0.85cm}

The discovery of $^{119}$Cd was reported in \textit{Decay of Cd$^{119}$ and In$^{119}$ Isomers} by Gleit and Coryell in 1961 \cite{Gle61b}. Deuterons accelerated to 14 MeV by the M.I.T. cyclotron bombarded natural uranium. $^{119}$Cd was produced by fission and identified by its $\beta$-decay following chemical separation. ``The true Cd half-life was calculated from a series of 14 experiments to be 2.7$\pm$0.3 min... The 2.7-min Cd appears a parent of both 18-min In and 2.0-min In and is tentatively identified as Cd$^{119m}$.'' The observed half-life could correspond to the isomeric state of $^{119}$Cd with a half-life of 2.20(2)~m; however it also could be the ground state which has a half-life of 2.69(2)~m. In the same paper, Gleit and Coryell incorrectly identified a half-life of 9.5~m for the ground state of $^{119}$Cd. A similar half-life of 10~m had previously also been incorrectly assigned to $^{119}$Cd \cite{Nus57}.

\subsection*{$^{120}$Cd}\vspace{-0.85cm}

The identification of $^{120}$Cd was reported in \textit{$^{120}$Cd and $^{122}$Cd} by Scheidemann and Hagebo in 1973 \cite{Sch73}. The CERN 600~MeV syncro-cyclotron was used to bombard a molten tin target with a 600 MeV proton beam. $^{120}$Cd, produced in spallation reactions, was separated and identified with the isotope separator on line (ISOLDE). ``The half-lives of $^{120}$Cd and $^{120}$In were measured on line by multianalysis of the spectrum or with the plastic detector and a scaler, and off line using a proportional-counter.'' The measured half-life of 50.80(21)~s is still the only half-life measurement of $^{120}$Cd. It should be mentioned that a half-life of $<$ 1~m was assigned to $^{120}$Cd already in 1961, however, no quantitative analysis was performed \cite{Gle61b}.

\subsection*{$^{121}$Cd}\vspace{-0.85cm}

Weiss reported the first observation of $^{121}$Cd in the 1965 article \textit{Near-Symmetric Fission-Identification and Yield of Cd-121} \cite{Wei65}. $^{121}$Cd was produced via thermal neutron fission of enriched $^{235}$U in the Vallecitos nuclear test reactor and the $\beta$-ray activity was measured on a gas-flow proportional $\beta$-ray counter. ``The curve which resulted from a plot of $^{121}$Sn activity as a function of separation time constituted a decay curve for $^{121}$Cd... The half-life of Cd$^{121}$ derived from these data by the method of least squares is 12.8$^{+0.4}_{-0.3}$ sec.'' This half-life is included in the weighted average of the currently accepted value of 13.5(3)~s.

\subsection*{$^{122}$Cd}\vspace{-0.85cm}

The identification of $^{122}$Cd was reported in \textit{$^{120}$Cd and $^{122}$Cd} by Scheidemann and Hagebo in 1973 \cite{Sch73}. The CERN 600~MeV syncro-cyclotron was used to bombard a molten tin target with a 600 MeV proton beam. $^{120}$Cd produced in spallation reactions was separated and identified with the isotope separator on line (ISOLDE). ``The half-life of $^{122}$Cd was measured on line with the plastic detector using the analyser as a multiscaler.'' The measured half-life of 5.78(9)~s is consistent with the presently accepted value of 5.24(3)~s. Scheidemann and Hagebo quote a previously measured half-life of 5.5(1)~s from an unpublished report by the OSIRIS collaboration \cite{Gra70}.

\subsection*{$^{123}$Cd}\vspace{-0.85cm}

In the 1983 paper \textit{Half-lives and emission probabilities of delayed neutron precursors $^{121-124}$Ag} Reeder \textit{et al.} described the discovery of $^{123}$Cd \cite{Ree83}. $^{123}$Cd was produced by thermal neutron-induced fission at Brookhaven National Laboratory and separated by the TRISTAN on-line isotope separator. ``From beta decay measurements a half-life of 2.07$\pm$0.03~s was found at mass 123 and assigned to the previously unknown $^{123}$Cd.'' This half-life is included in the weighted average of the currently accepted value of 2.10(2)~s. A previously reported half-life of 6~s \cite{Rud81} could not be confirmed.

\subsection*{$^{124}$Cd}\vspace{-0.85cm}

The discovery of $^{124}$Cd was reported in the 1974 article \textit{Levels and Transition Probabilities in $^{124}$In as Observed in the Decay of $^{124}$Cd} by Fogelberg \textit{et al.} \cite{Fog74}. $^{124}$Cd was produced by neutron induced fission at Studsvik, Sweden, and separated with the OSIRIS on-line facility. ``At mass 124, transitions showing half-lives of 0.9$\pm$0.2, 2.4$\pm$0.3 and 3.2$\pm$0.3~s were found, and were identified as being due to the decay of $^{124}$Cd and a high and a low spin isomer of $^{124}$In, respectively.''  The half-life of 0.9(2)~s is consistent with the currently accepted value of 1.25(2)~s.

\subsection*{$^{125}$Cd}\vspace{-0.85cm}

The article \textit{Measurements of Absolute $\gamma$-ray Intensities in the Decays of Very Neutron Rich Isotopes of Cd and In} by Gokturk \textit{et al.} reported the discovery of $^{125}$Cd in 1986 \cite{Gok86}. $^{125}$Cd was formed by neutron induced fission of a $^{235}$U target at the OSIRIS ISOL-facility at Studsvik, Sweden. ``Four previously unknown or little known isotopes of Cd are reliably characterized for the first time.'' The half-life $^{125}$Cd was measured to be 0.75(4)~s which is consistent with the currently accepted value of 0.65(2)~s. A previously reported half-life of 12.2~s \cite{Rud81} could not be confirmed.

\subsection*{$^{126}$Cd}\vspace{-0.85cm}

Gartner and Hill reported the first observation of $^{126}$Cd in the 1978 article \textit{Decay of Mass separated $^{126}$Cd} \cite{Gar78}. $^{126}$Cd was produced in neutron induced fission at the Ames Laboratory Research Reactor and separated with the Tristan on-line isotope-separator. ``$\gamma$ rays from the decay of $^{126}$Cd were observed using Ge(Li) and LEPS detectors located near the point of beam deposition... We propose therefore that the half-life of $^{126}$Cd is 0.506$\pm$0.015 sec.'' This half-life is included in the weighted average of the currently accepted value of 0.515(17)~s. Gartner and Hill quote a previously measured half-life of 0.53~s from an unpublished report by the OSIRIS collaboration \cite{Gra74}.

\subsection*{$^{127-129}$Cd}\vspace{-0.85cm}

The article \textit{Measurements of Absolute $\gamma$-ray Intensities in the Decays of Very Neutron Rich Isotopes of Cd and In} by Gokturk \textit{et al.} reported the discovery of $^{127}$Cd, $^{128}$Cd, and $^{129}$Cd in 1986 \cite{Gok86}.  These cadmium isotopes were formed by neutron induced fission of $^{235}$U target at the OSIRIS ISOL-facility at Studsvik, Sweden. ``Four previously unknown or little known isotopes of Cd are reliably characterized for the first time.'' The half-lives reported for $^{127}$Cd (0.43(3)~s) and $^{128}$Cd (0.34(3)~s) are both included in the weighted average of the currently accepted values of 0.37(7)~s and 0.28(4)~s, respectively. The half-life of 0.27(4)~s is the only measured value for $^{129}$Cd listed in the ENSDF database and is consistent with the most recent result of 0.242(8)~s quoted in the NUBASE evaluation. A previously reported half-life of 0.9~s for $^{128}$Cd \cite{Rud81} could not be confirmed.

\subsection*{$^{130}$Cd}\vspace{-0.85cm}

The discovery of $^{130}$Cd was reported in 1986 by Kratz \textit{et al.} in \textit{The Beta-Decay Half-Life of $^{130}$Cd and its Importance for Astrophysical r-Process Scenarios} \cite{Kra86}. A target of Uranium carbide was bombarded with 600 MeV protons from the CERN synchrocyclotron. $^{130}$Cd was produced in spallation reactions and subsequently separated and identified with the ISOLDE on-line mass separator. ``From the $\beta$dn-growth curve [$\beta$-delayed-neutron] the T$_{1/2}$ of $^{130}$Cd was determined to (195 $\pm$ 35) ms.'' This half-life is consistent with the currently accepted value of 162(7)~ms.

\subsection*{$^{131,132}$Cd}\vspace{-0.85cm}

$^{131}$Cd and $^{132}$Cd were first discovered in 2000 by Hannawald as reported in the article \textit{Selective Laser Ionization of very neutron-rich cadmium isotopes: Decay properties of $^{131}$Cd and $^{132}$Cd} \cite{Han00}. A uranium-carbide/graphite target was bombarded by 1~GeV protons at CERN. $^{131}$Cd and $^{132}$Cd were separated and identified using a resonance-ionization laser ion-source at ISOLDE. ``For the new $N$ = 83 isotope $^{131}$Cd a half-life of $T_{1/2}$ = (68$\pm$3) ms has been obtained, which is unanticipatedly short when compared to current model predictions... A somewhat similar picture arises for the new $N$ = 84 isotope $^{132}$Cd, where the experimental half-life of $T_{1/2}$ = (97$\pm$10) ms is again considerably shorter than the predicted $T_{1/2}$(GT)=633 ms...'' These half-lives of 68(3)~ms and 97(10)~ms determined for $^{131}$Cd and $^{132}$Cd, respectively, are the currently accepted values.

\section{Summary}
The discoveries of the known cadmium isotopes have been compiled and the methods of their production discussed. The first measured half-lives of several isotopes ($^{101}$Cd, $^{102}$Cd, $^{119}$Cd, $^{123}$Cd, $^{125}$Cd, and $^{128}$Cd) were incorrect. The half-lives of $^{107}$Cd and $^{109}$Cd were first reported without a definite mass assignment. In addition, the initial assignments of $^{107}$Cd and $^{109}$Cd, as well as $^{115}$Cd, and $^{117}$Cd, were reversed. The half-lives of $^{118}$Cd, $^{122}$Cd, and $^{126}$Cd were first mentioned in a conference proceeding or unpublished reports several years prior to publication in refereed journals. Finally, $^{115}$Cd was initial incorrectly reported as being a stable isotope.

\ack
This work was supported by the National Science Foundation under grants No. PHY06-06007 (NSCL) and PHY07-54541 (REU).

%%% Here we use thebibliography environment to produce the reference list,
%%% but you can use BibTeX as well:
%\bibliography{tmpadnd}

\newpage

\section*{EXPLANATION OF TABLE}\label{sec.eot}
\addcontentsline{toc}{section}{EXPLANATION OF TABLE}

\renewcommand{\arraystretch}{1.0}

\begin{tabular*}{0.95\textwidth}{@{}@{\extracolsep{\fill}}lp{5.5in}@{}}
\textbf{TABLE I.}
	& \textbf{Discovery of Cadmium Isotopes }\\
%\multicolumn{2}{p{0.95\textwidth}}{(Throughout this table,
%	italicized numbers refer to derived values)}
\\

Isotope &  Cadmium isotope \\
Author & First author of refereed publication \\
Journal & Journal of publication \\
Ref. &  Reference  \\
Method & Production method used in the discovery: \\
    & FE: fusion evaporation \\
    & LP: light-particle reactions (including neutrons) \\
    & NC: neutron capture \\
    & MS: mass spectroscopy \\
    & SP: spallation \\
    & PF: projectile fragmentation or fission \\
    & NF: neutron induced fission \\
    & CPF: charged particle induced fission \\
Laboratory &  Laboratory where the experiment was performed\\
Country &  Country of laboratory\\
Year & Year of discovery  \\
\end{tabular*}
\label{tableI}

\newpage
\datatables

%% If the table is to span over the whole text width, we set:
\setlength{\LTleft}{0pt}
\setlength{\LTright}{0pt}

% To avoid ``Overfull \hboxes...'' decrease the intercolumn spacing:

\setlength{\tabcolsep}{0.5\tabcolsep}

\renewcommand{\arraystretch}{1.0}

%%\footnotesize % we need to squeeze the font size a lot!

\begin{longtable}[c]{%
@{}@{\extracolsep{\fill}}r@{\hspace{5\tabcolsep}} llllllll@{}}
\caption[Discovery of Cadmium Isotopes]%
{Discovery of Cadmium isotopes}\\[0pt]
\caption*{\small{See page \pageref{tableI} for Explanation of Tables}}\\
\hline
\\[100pt]
\multicolumn{8}{c}{\textit{This space intentionally left blank}}\\
\endfirsthead
Isotope & First Author & Journal & Ref. & Method & Laboratory & Country & Year \\

$^{96}$Cd & D. Bazin & Phys. Rev. Lett. & Baz08 & PF & Michigan State & USA &2008 \\
$^{97}$Cd & T. Elmroth & Nucl. Phys. A & Elm78 & SP & CERN & Switzerland &1978 \\
$^{98}$Cd & T. Elmroth & Nucl. Phys. A & Elm78 & SP & CERN & Switzerland &1978 \\
$^{99}$Cd & T. Elmroth & Nucl. Phys. A & Elm78 & SP & CERN & Switzerland &1978 \\
$^{100}$Cd & D.J. Hnatowich & J. Inorg. Nucl. Chem. & Hna70 & SP & CERN & Switzerland &1970 \\
$^{101}$Cd & P.G. Hansen & Phys. Lett. B & Han69 & SP & CERN & Switzerland &1969 \\
$^{102}$Cd & P.G. Hansen & Phys. Lett. B & Han69 & SP & CERN & Switzerland &1969 \\
$^{103}$Cd & I.L. Preiss & Nucl. Phys. & Pre60 & FE & Yale & USA &1960 \\
$^{104}$Cd & F.A. Johnson & Can. J. Phys. & Joh55 & LP & McGill & Canada &1955 \\
$^{105}$Cd & J.R. Gum & Phys. Rev. & Gum50 & LP & Ohio State & USA &1950 \\
$^{106}$Cd & F.W. Aston & Proc. Roy. Soc. & Ast35 & MS & Cambridge & UK &1935 \\
$^{107}$Cd & A.C. Helmholz & Phys. Rev. & Hel46 & LP & Berkeley & USA &1946 \\
$^{108}$Cd & F.W. Aston & Proc. Roy. Soc. & Ast35 & MS & Cambridge & UK &1935 \\
$^{109}$Cd & M. Lindner & Phys. Rev. & Lin50 & SP & Berkeley & USA &1950 \\
$^{110}$Cd & F.W. Aston & Phil. Mag. & Ast25 & MS & Cambridge & UK &1925 \\
$^{111}$Cd & F.W. Aston & Phil. Mag. & Ast25 & MS & Cambridge & UK &1925 \\
$^{112}$Cd & F.W. Aston & Phil. Mag. & Ast25 & MS & Cambridge & UK &1925 \\
$^{113}$Cd & F.W. Aston & Phil. Mag. & Ast25 & MS & Cambridge & UK &1925 \\
$^{114}$Cd & F.W. Aston & Phil. Mag. & Ast25 & MS & Cambridge & UK &1925 \\
$^{115}$Cd & M. Goldhaber & Phys. Rev. & Gol39 & LP & Cambridge & UK &1939 \\
$^{116}$Cd & F.W. Aston & Phil. Mag. & Ast25 & MS & Cambridge & UK &1925 \\
$^{117}$Cd & M. Goldhaber & Phys. Rev. & Gol39 & NC & Cambridge & UK &1939 \\
$^{118}$Cd & C.E. Gleit & Phys. Rev. & Gle61a & CPF & MIT & USA &1961 \\
$^{119}$Cd & C.E. Gleit & Phys. Rev. & Gle61b & CPF & MIT & USA &1961 \\
$^{120}$Cd & O. Scheidemann & J. Inorg. Nucl. Chem. & Sch73 & SP & CERN & Switzerland &1973 \\
$^{121}$Cd & H.V. Weiss & Phys. Rev. & Wei65 & NF & U.S. Naval Rad. Def. Lab. & USA &1965 \\
$^{122}$Cd & O. Scheidemann & J. Inorg. Nucl. Chem. & Sch73 & SP & CERN & Switzerland &1973 \\
$^{123}$Cd & P.L. Reeder& Phys. Rev. C & Ree83 & NF & Brookhaven & USA &1983 \\
$^{124}$Cd & B. Fogelberg & Nucl. Phys. A & Fog74 & NF & Studsvik & Sweden &1974 \\
$^{125}$Cd & H. Gokturk & Z. Phys. A & Gok86 & NF & Studsvik & Sweden &1986 \\
$^{126}$Cd & M.L. Gartner & Phys. Rev. C & Gar78 & NF & Ames & USA &1978 \\
$^{127}$Cd & H. Gokturk & Z. Phys. A & Gok86 & NF & Studsvik & Sweden &1986 \\
$^{128}$Cd & H. Gokturk & Z. Phys. A & Gok86 & NF & Studsvik & Sweden &1986 \\
$^{129}$Cd & H. Gokturk & Z. Phys. A & Gok86 & NF & Studsvik & Sweden &1986 \\
$^{130}$Cd & K.L. Kratz & Z. Phys. A & Kra86 & SP & CERN & Switzerland &1986 \\
$^{131}$Cd & M. Hannawald & Phys. Rev. C & Han00 & SP & CERN & Switzerland &2000 \\
$^{132}$Cd & M. Hannawald & Phys. Rev. C & Han00 & SP & CERN & Switzerland &2000 \\

\end{longtable}

\newpage

%% A long reference list can be squeezed using:
%\newcommand{\bibfont}{\footnotesize}

\normalsize

\begin{theDTbibliography}{1956He83}

\bibitem[Ast25]{Ast25t} F.W. Aston, Phil. Mag. {\bf 49}, 1191 (1925)
\bibitem[Ast35]{Ast35t} F.W. Aston, Proc. Roy. Soc. A {\bf 149}, 396 (1935)
\bibitem[Baz08]{Baz08t} D. Bazin, F. Montes, A. Becerril, G. Lorusso, A. Amthor, T. Baumann, H. Crawford, A. Estrade, A. Gade, T. Ginter, C.J. Guess, M. Hausmann, G.W. Hitt, P. Mantica, M. Matos, R. Meharchand, K. Minamisono, G. Perdikakis, J. Pereira, J. Pinter, M. Portillo, H. Schatz, K. Smith, J. Stoker, A. Stolz, and R.G.T. Zegers, Phys. Rev. Lett. {\bf 101}, 252501 (2008)
\bibitem[Elm78]{Elm78t} T. Elmroth, E. Hagberg, P.G. Hansen, J.C. Hardy, B. Jonson, H.L. Ravn and P. Tidemand-Petersson, Nucl. Phys. A {\bf304}, 493 (1978)
\bibitem[Fog74]{Fog74t} B. Fogelberg, T. Nagarajan and B. Grapengiesser, Nucl. Phys. {\bf A230}, 214 (1974)
\bibitem[Gar78]{Gar78t} M.L. Gartner and J.C. Hill, Phys. Rev. C {\bf 18}, 1463 (1978)
\bibitem[Gle61a]{Gle61at} C.E. Gleit and C.D. Coryell, Phys. Rev. {\bf 122}, 229 (1961)
\bibitem[Gle61b]{Gle61bt} C.E. Gleit and C.D. Coryell, Phys. Rev. {\bf 124}, 1914 (1961)
\bibitem[Gok86]{Gok86t} H. Gokturk, B. Ekstrom, E. Lund, and B. Fogelberg, Z. Phys. A {\bf 324}, 117 (1986)
\bibitem[Gol39]{Gol39t} M. Goldhaber, R.D. Hill, and L. Szilard, Phys. Rev. {\bf 55}, 47 (1939)
\bibitem[Gum50]{Gum50t} J.R. Gum and M.L. Pool, Phys. Rev. {\bf 80}, 315 (1950)
\bibitem[Han69]{Han69t} P.G. Hansen, P. Hornshoj, H.L. Nielsen, K. Wilsky, H. Kugler, G. Astner, E. Hagebo, J. Hudis, A. Kjelberg, F. M\"unnich, P. Patzelt, M. Alpsten, G. Andersson, Aa. Appelqvist, B. Bengtsson, R.A. Naumann, O.B. Nielsen, E. Beck, R. Foucher, P. Husson, J. Jastrzebski, A. Johnson, J. Alstad, T. Jahnsen, A.C. Pappas and T. Tunaal, R. Henck, P. Siffert, and G. Rudstam, Phys. Lett. {\bf 28B}, 415 (1969)
\bibitem[Han00]{Han00t} M. Hannawald, K.L. Kratz, B. Pfeiffer, W.B. Walters, V.N. Fedoseyev, V.I. Mishin, W.F. Mueller, H. Schatz, J. Van Roosbroeck, U. K\"oster, V. Sebastian, H.L. Ravn, and the ISOLDE Collaboration, Phys. Rev. C {\bf 62}, 054301 (2000)
\bibitem[Hel46]{Hel46t} A.C. Helmholz, Phys. Rev. {\bf 70}, 982 (1946)
\bibitem[Hna70]{Hna70t} D.J. Hnatowich, E. Hagebo, A. Kjelberg, R. Mohr, and P. Patzelt, J. Inorg. Nucl. Chem. {\bf 32}, 3137 (1970)
\bibitem[Joh55]{Joh55t} F.A. Johnson, Can. J. Phys. {\bf 33}, 841 (1955)
\bibitem[Kra86]{Kra86t} K.L. Kratz, H. Gabelmann, W. Hillebrandt, B. Pfeiffer, K. Schlosser, and F.K. Thielemann, Z. Phys. A {\bf 325}, 489 (1986)
\bibitem[Lin50]{Lin50t} M. Lindner and I. Perlman, Phys. Rev. {\bf 78}, 499 (1950)
\bibitem[Pre60]{Pre60t} I.L. Preiss, P.J. Estrup and R. Wolfgang, Nucl. Phys. {\bf 18}, 624 (1960)
\bibitem[Ree83]{Ree83t} P. L. Reeder, and R.A. Warner, R.L. Gill, Phys. Rev. {\bf C27}, 3002 (1983)
\bibitem[Sch73]{Sch73t} 0. Scheidemann and E. Hagebo, J. Inorg. Nucl. Chem. {\bf 35}, 3055 (1973)
\bibitem[Wei65]{Wei65t} H. V. Weiss, Phys. Rev. {\bf 139}, B 304 (1965)

\end{theDTbibliography}

\end{document}